\documentstyle[12pt,epsf]{article}
\def\beq{\begin{equation}}
\def\eeq{\end{equation}}
\def\barr{\begin{eqnarray}}
\def\earr{\end{eqnarray}}
\def\b{\bigskip}
\def\n{\noindent}
\def\m{\medskip}

\begin{document}

\title{CLASSIFICATION OF NON-ABELIAN CHERN-SIMONS VORTICES\footnote{Talk
presented at the {\it $XXII^{nd}$ International Conference on Differential
Geometric Methods in Theoretical Physics}, Ixtapa (Mexico), September 1993.}}
\author{Gerald V. Dunne\\
Department of Physics\\
University of Connecticut\\2152 Hillside Road\\
Storrs, CT 06269 USA\\
dunne@hep.phys.uconn.edu}

\maketitle

\begin{abstract}
The two-dimensional self-dual Chern-Simons equations are equivalent to the
conditions for static, zero-energy vortex-like solutions of the (2+1)
dimensional gauged nonlinear Schr\"odinger equation with Chern-Simons
matter-gauge coupling. The finite charge vacuum states in the Chern-Simons
theory are shown to be gauge equivalent to the finite action solutions to the
two-dimensional chiral model (or harmonic map) equations. The Uhlenbeck-Wood
classification of such harmonic maps into the unitary groups thereby leads to a
complete classification of the vacuum states of the Chern-Simons model. This
construction also leads to an interesting new relationship between $SU(N)$ Toda
theories and the $SU(N)$ chiral model.

\end{abstract}
\b\b

The study of the nonlinear Schr\"odinger equation in $2+1$-dimensional
space-time is partly motivated by the well-known success of the
$1+1$-dimensional nonlinear Schr\"odinger equation. Here we consider a {\it
gauged} nonlinear Schr\"odinger equation in which we have not only the
nonlinear potential term for the matter fields, but also we have a coupling of
the matter fields to the gauge fields. Furthermore, this matter-gauge dynamics
is chosen to be of the Chern-Simons form rather than the conventional
Yang-Mills form. With this choice, the nonlinear term in the Schr\"odinger
equation may also be viewed as a Pauli interaction, due to the Chern-Simons
relation between the magnetic field and the charge density.

The theory with an Abelian gauge field was analyzed by Jackiw and Pi
\cite{Jack} who found static, zero energy solutions which arise from a
two-dimensional notion of self-duality. The static, self-dual matter density
satisfies the Liouville equation, which is known to be integrable \cite{Liou}.
The gauged nonlinear Schr\"odinger equation with {\it non-Abelian} Chern-Simons
matter-gauge dynamics has also been considered \cite{Gross,Dun1,Dun2}, and once
again static, zero energy solutions (referred to as "{\it self-dual
Chern-Simons vortices}") have been found to arise from an analogous, but much
richer, two-dimensional self-duality condition. These two-dimensional
self-duality equations are formally integrable and in special cases they reduce
to the classical and affine Toda equations, both known integrable systems of
nonlinear partial differential equations \cite{Kost,Lezn}.

Here, I classify all finite charge solutions to the self-dual Chern-Simons
equations by first showing that the self-duality equations are equivalent to
the classical equations of motion of the Euclidean two-dimensional chiral model
(also known as the harmonic map equations), and then using a deep
classification theorem due to K.~Uhlenbeck \cite{KUhl} which classifies all
$U(N)$ and $SU(N)$ chiral model solutions with finite chiral model action. The
chiral model action is in fact proportional to the net gauge invariant {\it
charge} $Q$ in the matter-Chern-Simons system, and so the classification of all
finite charge solutions is complete. I also present the explicit "{\it uniton}"
decomposition of a special class of solutions to the $SU(N)$ chiral model
equations which have the remarkable property that when the matter density for
these solutions is diagonalized, it satisfies the classical $SU(N)$ Toda
equations. Such a direct correspondence between the Toda equations and the
chiral model equations is surprising.

The $2+1$-dimensional nonlinear Schr\"odinger equation reads \footnote{Note
that there is a typographical error in this equation in \cite{Dun2}.}
\beq
i D_0 \Psi=-{1\over 2}\vec{D}^2\Psi +{1\over\kappa}[~[\Psi,\Psi^{\dagger}],
\Psi]\
\label{Schr}
\eeq
where the covariant derivative is $D_\mu \equiv\partial_\mu +[A_\mu ,~~]$, and
both the gauge potential $A_\mu$ and the matter field $\Psi$ are Lie algebra
valued: $A_\mu =A_\mu ^a T^a$, $\Psi=\Psi^a T^a$. The main results of this
paper are for the Lie algebra of $SU(N)$, but the formulation generalizes
straightforwardly to any simple Lie algebra (the noncompact case has been
studied in \cite{Cang}). The matter and gauge fields are coupled dynamically by
the Chern-Simons equation
\beq
F_{\mu \nu} = {i\over\kappa} \epsilon_{\mu \nu \rho} J^\rho\
\label{Cher}
\eeq
where $F_{\mu \nu}=\partial_\mu A_\nu -\partial_\nu A_\mu +[A_\mu,A_\nu]$ is
the gauge curvature, $\kappa$ is a coupling constant and $J^\rho$ is the
covariantly conserved ($D_\mu J^\mu =0$) nonrelativistic matter current
\barr
J^0 &=&[\Psi^{\dagger},\Psi]\nonumber\\
J^i&=&-{i\over 2}\left([\Psi^{\dagger},D_i \Psi] - [(D_i \Psi)^{\dagger},
\Psi]\right)\
\label{curr}
\earr
The Schr\"odinger equation (\ref{Schr}) and the Chern-Simons equation
(\ref{Cher}) are invariant under the gauge transformation
\barr
\Psi&\to&g^{-1} \Psi g\nonumber\\
A_\mu&\to&g^{-1}A_\mu g +g^{-1}\partial_\mu g\
\label{gaug}
\earr
where $g\in SU(N)$.

In \cite{Dun1,Dun2} it has been shown that the minimum (in fact {\it zero})
energy solutions to (\ref{Schr},\ref{Cher}) are given by the {\it self-dual
Ansatz}
\beq
D_- \Psi=0\
\label{self}
\eeq
combined with the remaining Chern-Simons equation
\beq
\partial_- A_+ -\partial_+ A_- +[A_-,A_+] = {2\over\kappa}~[\Psi^{\dagger},
\Psi]\
\label{dual}
\eeq
Here $A_\pm =A_1\pm iA_2$, $D_\pm =D_1 \pm iD_2$ and with antihermitean Lie
algebra generators we have $A_\pm =-(A_\mp)^{\dag}$. Equations
(\ref{self},\ref{dual}) are collectively referred to as the {\it self-dual
Chern-Simons equations}. The  self-dual solutions provide {\it static}
solutions to the gauged nonlinear Schr\"odinger equation, as can be seen from a
Hamiltonian formulation \cite{Dun1}. Alternatively, this follows directly from
the equations of motion (\ref{Schr},\ref{Cher}). To see this, note that if
$D_{-}\Psi=0$, then the currents take the simple form
\barr
J^{+}&\equiv&J^1 +i J^2\nonumber\\
&=&-{i\over 2}[\Psi^{\dagger}, D_+ \Psi]
\label{current}
\earr
It then follows from the Chern-Simons equation (\ref{Cher}) that $A_0={i\over
2\kappa} [\Psi^{\dagger}, \Psi]$. The identity
\barr
\vec{D}^2 \Psi&\equiv &D_+ D_- \Psi + i [F_{12}, \Psi]\nonumber\\
&=&D_+ D_- \Psi - {1\over \kappa} [ [\Psi^{\dagger}, \Psi], \Psi]
\label{ident}
\earr
then implies that the Schr\"odinger equation reduces to
\beq
i\partial_0 \Psi= -{1\over 2} D_+ D_- \Psi =0
\label{zero}
\eeq
In fact, owing to a remarkable dynamical $SO(2,1)$ symmetry of the gauged
nonlinear Chern-Simons-Schr\"odinger equations (\ref{Schr},\ref{Cher}), it is
possible to show that the implication holds in the reverse direction: {\it all}
static solutions are self-dual \cite{Dun1}.

Before classifying the general solutions to the self-dual Chern-Simons
equations, it is instructive to consider certain special cases in which
simplifying algebraic {\it Ans\"atze} for the fields reduce
(\ref{self},\ref{dual}) to familiar integrable nonlinear equations. First,
choose the fields to have the following Lie algebra decomposition
\beq
A_i = \sum_{\alpha}A_i^{\alpha} H_\alpha ~~~~~~~~~~~~~~~~~~\Psi = \sum_{\alpha}
\psi^{\alpha} E_{\alpha}\
\label{ans1}
\eeq
where the sum is over all positive, simple roots $\alpha$ of the Lie algebra,
and $H_\alpha$ and $E_\alpha$ are the Cartan subalgebra and step operator
generators (respectively) in the Chevalley basis \cite{Hump}. Then the
self-dual Chern-Simons equations (\ref{self},\ref{dual}) combine to yield the
classical Toda equations
\beq
\nabla ^2 {\rm log} \rho_\alpha = -{2\over\kappa} K_{\alpha \beta} \rho_\beta\
\label{toda}
\eeq
where $\rho_\alpha \equiv |\psi^\alpha |^2$, and $K_{\alpha \beta}$ is the
classical Cartan matrix for the Lie algebra. For $SU(2)$, (\ref{toda}) becomes
the Liouville equation $\nabla ^2 {\rm log}\rho = -{4\over\kappa} \rho$, which
Liouville showed to be integrable and indeed "solvable" \cite{Liou} - in the
sense that the general real solution can be expressed in terms of a single
holomorphic function $f=f(x^-)$:
\beq
\rho={\kappa\over 2} \nabla ^2 {\rm log} \left(1+f(x^-) \bar{f}(x^+)\right)\
\label{liou}
\eeq
Kostant \cite{Kost}, and Leznov and Saveliev \cite{Lezn} have shown that the
classical Toda equations (\ref{toda}) are similarly integrable, with the
general real solutions for $\rho_\alpha$ being expressible in terms of $r$
arbitrary holomorphic functions, where $r$ is the rank of the algebra. For
$SU(N)$ it is possible to adapt the Kostant-Leznov-Saveliev solutions to a
simpler form more reminiscent of the Liouville solution (\ref{liou}):
\beq
\rho_\alpha={\kappa\over 2} \nabla ^2 {\rm log~det}
\left(M_{\alpha}^{\dag}(x^+) M_{\alpha}(x^-)\right)\
\label{sunn}
\eeq
where $M_\alpha$ is the $N{\times}\alpha$ {\it rectangular} matrix $M_\alpha =
(u~\partial_-u~\partial_-^2 u~\dots \partial_-^{\alpha-1} u)$, with $u$ being
an $N$-component column vector
\beq
u=\left(\matrix{1\cr f_1(x^-)\cr f_2(x^-) \cr \vdots\cr
f_{N-1}(x^-)\cr}\right)\
\label{uvec}
\eeq

\begin{figure}[t]
\centering    \vspace{1cm}
    \epsffile{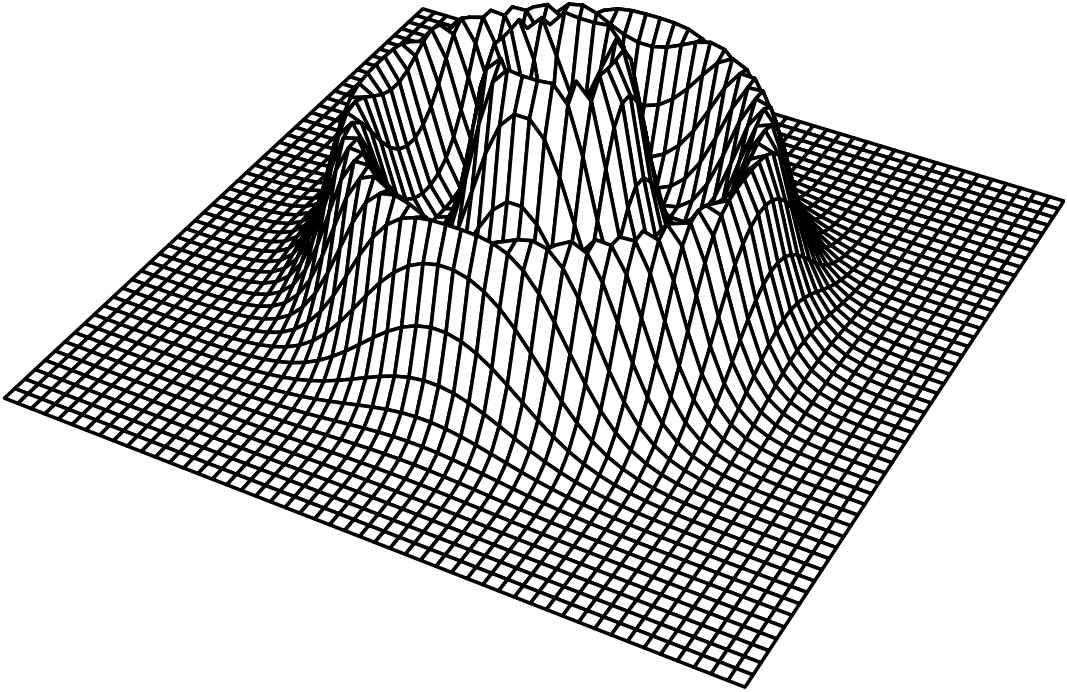}
    \caption{A plot of the nonAbelian charge density $\rho_1$ for a radially
symmetric SU(3) Toda-type vortex solution (\protect{\ref{sunn}}) to the
self-dual Chern-Simons equations (\protect{\ref{self}},\protect{\ref{dual}}).
For a radially symmetric solution, the functions $f_\alpha(x^-)$ appearing in
(\protect{\ref{uvec}}) are chosen to be powers of $x^-$.}
    \label{corrplot}\end{figure}

For a radially symmetric $SU(3)$ example see Figure 1. An alternative,
extended, {\it Ans\"atz} involves the matter field choice
\beq
\Psi = \sum_{\alpha} \psi^{\alpha} E_{\alpha} +\psi^M E_{-M}\
\label{ans2}
\eeq
where $E_{-M}$ is the step operator corresponding to minus the maximal root.
With the gauge field still as in (\ref{ans1}), the self-dual Chern-Simons
equations then combine to give the {\it affine} Toda equations
\beq
\nabla ^2 {\rm log} \rho_a = -{2\over\kappa} \tilde{K}_{a b} \rho_b\
\label{afftoda}
\eeq
where $\tilde{K}$ is the affine Cartan matrix. These affine Toda equations are
also known to be integrable.

Having considered some special cases of solutions to the self-dual Chern-Simons
equations, we now consider the general solutions by first making a gauge
transformation to convert the equations (\ref{self},\ref{dual}) into the {\it
single} equation
\beq
\partial_- \chi = [\chi^{\dag}, \chi ]\
\label{sing}
\eeq
where $\chi$ is the gauge transformed matter field $\chi = \sqrt{{2\over
\kappa}} g \Psi g^{-1}$. The existence of such a gauge transformation $g^{-1}$
follows from the following zero-curvature formulation of the self-dual
Chern-Simons equations \cite{Dun1,Dun2}. Define
\beq
{\cal{A}}_+ \equiv A_+ - \sqrt{{2\over \kappa}} \Psi ,~~~~~~~~~~~~~{\cal{A}}_-
\equiv A_- + \sqrt{{2\over \kappa}} \Psi^{\dag}\
\label{curl}
\eeq
Then the self-dual Chern-Simons equations imply that the gauge curvature
associated with ${\cal{A}}_\pm$ vanishes: $\partial_-{\cal{A}}_+ -\partial_+
{\cal{A}}_- +[{\cal{A}}_-, {\cal{A}}_+] = 0$. Therefore, locally, one can write
$\cal{A}_\pm$ as pure gauge
\beq
{\cal{A}}_\pm  = g^{-1} \partial_\pm g \
\label{pure}
\eeq
for some $g\in SU(N)$. Gauge transforming the self-dual Chern-Simons equations
(\ref{self},\ref{dual}) with this group element $g^{-1}$ leads to the single
equation (\ref{sing}).

Equation (\ref{sing}) can be converted into the chiral model equation by
defining $\chi={1\over 2} h^{-1}\partial_+ h$ for some $h\in SU(N)$ (the fact
that it is possible to write $\chi$ in this manner is a consequence of
(\ref{sing})). The chiral model equation \cite{Zakr} reads:
\beq
\partial_+ (h^{-1} \partial_- h) + \partial_- (h^{-1} \partial_+ h) = 0\
\label{chir}
\eeq
Given any solution $h$ of the chiral model equations, or alternatively any
solution $\chi$ of (\ref{sing}), we automatically obtain a solution of the
original self-dual Chern-Simons equations:
\beq
\Psi^{(0)} = \sqrt{{\kappa\over 2}}\chi ,~~~~~~~A_+ ^{(0)}=\chi ,~~~~~~~A_-
^{(0)}=-\chi ^{\dag} .\
\label{spec}
\eeq

The {\it global} condition which permits the classification of solutions to the
chiral model equation (\ref{chir}) is that the chiral model "action functional"
(also referred to in the literature as the "energy functional")
\beq
{\cal{E}}[h] = -{1\over 2} \int d^2x~tr(h^{-1}\partial_- h h^{-1} \partial_+
h)\
\label{ener}
\eeq
be {\it finite}. This finiteness condition is directly relevant in the related
matter-Chern-Simons system because ${\cal{E}}[h] = 2\int d^2 x~tr(\chi
\chi^{\dag}) = {4\over\kappa} \int d^2x~tr(\Psi \Psi^{\dag})={4\over\kappa} Q$
where $Q$ is the net gauge invariant matter charge. As well as being physically
significant, this finiteness condition is mathematically crucial because it
permits the chiral model solutions on ${\bf R}^2$ to be classified by conformal
compactification to the sphere $S^2$ \cite{KUhl,Ward}.
\b

\n{\bf THEOREM} (K.~Uhlenbeck \cite{KUhl}; see also J.~C.~Wood \cite{Wood}):
{\it Every finite action solution h of the SU(N) chiral model equation
(\ref{chir}) may be uniquely factorized as a product of "uniton" factors}
\beq
h=\pm h_0 \prod_{i=1}^m (2 p_i -1)\
\label{fact}
\eeq
{\it where:}

\n{\it a) $h_0 \in SU(N)$ is constant;}

\n{\it b) each $p_i$ is a Hermitean projector ($p_i^{\dag}=p_i$ and
$p_i^2=p_i$);}

\n{\it c) defining $h_j=h_0 \prod_{i=1}^j (2p_i-1)$, the following linear
relations must hold:}
\barr
(1-p_i) \left(\partial_+ + {1\over 2} h_{i-1}^{-1} \partial_+
h_{i-1}\right)~p_i&=&0\nonumber\cr
(1-p_i)~h_{i-1}^{-1} \partial_- h_{i-1}~p_i &=&0\cr
\label{linr}
\earr
\n{\it d) $m\leq N-1$.}

\smallskip

\n The $\pm$ sign in (\ref{fact}) has been inserted to allow for the fact that
Uhlenbeck and Wood actually considered $U(N)$ rather than $SU(N)$.

An important implication of this theorem is that for $SU(2)$ {\it all} finite
action solutions of the chiral model have the "single uniton" form
\beq
h=-h_0 (2p-1)\
\label{unit}
\eeq
where $p$ is a holomorphic projector satisfying
\beq
(1-p)~\partial_+ p=0\
\label{holo}
\eeq
These solutions are essentially the $CP^1$ model solutions of Din and
Zakrzewski \cite{DinZ,Zakr}.

At this point, it is not at all obvious how these types of solutions to the
chiral model equations (and therefore by (\ref{spec}) of the self-dual
Chern-Simons equations) are related to the special Toda-type solutions
discussed previously. The key observation is that while the algebraic {\it
Ans\"atze} (\ref{ans1},\ref{ans2}) each lead to a non-Abelian charge density
$\rho=[\Psi^{\dag}, \Psi]$ which is {\it diagonal}, the chiral model solutions
(\ref{spec}) have charge density $\rho^{(0)} = {\kappa\over 2}[\chi^{\dag},
\chi]$ which need not be diagonal. However, $\rho$ is always hermitean, and so
it can be diagonalized by a gauge transformation. It is still an algebraically
nontrivial task to implement this diagonalization, but this is achieved below
for the solutions of $SU(N)$ Toda type.

It is instructive to illustrate this procedure with the $SU(2)$ case first.
Since $p^2=p$, the holomorphic projector condition (\ref{holo}) is equivalent
to the condition $\partial_+p~p=0$. All such projectors may be written as
\beq
p=M (M^{\dag} M)^{-1} M^{\dag}\
\label{proj}
\eeq
where $M(x^-)$ is any rectangular matrix depending only on the $x^-$ variable.
For $SU(2)$ we can only project onto a {\it line} in ${\bf C}^2$, so we take
\beq
M=\left(\matrix{1\cr f(x^-)\cr}\right)\
\label{mvec}
\eeq
This then leads to
\beq
p={1\over 1+f \bar{f}} \left(\matrix{1&\bar{f}\cr f& f \bar{f}\cr} \right)
{}~~~~~~~~~~~~\chi = \partial_+p = {f\partial_+ \bar{f} \over (1+f\bar{f})^2}
\left(\matrix{-1&1/f \cr-f&1\cr}\right)\
\label{pchi}
\eeq
The corresponding matter density is
\beq
[\chi^{\dag}, \chi] = -{\partial_+ \bar{f} \partial_- f\over (1+f\bar{f})^3}
\left(\matrix{1-f\bar{f}&2\bar{f} \cr 2f&-1+f\bar{f} \cr}\right)\
\label{dens}
\eeq
which may be diagonalized by the unitary matrix
\barr
g&=&{1\over \sqrt{1+f\bar{f}}} \left(\matrix{-\bar{f}&1\cr
1&f\cr}\right)\nonumber\\
g^{-1}[\chi^{\dag}, \chi] g&=&\partial_+ \partial_- {\rm log~det}(M^{\dag}
M)~\left(\matrix{1&0\cr 0&-1\cr}\right)\
\label{diag}
\earr
This is precisely Liouville's solution (\ref{liou}) to the classical $SU(2)$
Toda equation.

For the $SU(N)$ chiral model with $N\geq3$ it becomes increasingly difficult to
describe systematically all possible uniton factorizations consistent with the
linear relations listed in Uhlenbeck's theorem, but Wood \cite{Wood} has given
an explicit construction and parametrization of all $SU(N)$ solutions in terms
of sequences of Grassmannian factors.

Another useful result from the chiral model literature is due to Valli:
\m

\n{\bf THEOREM} (G.~Valli \cite{Vall}): {\it Let $h$ be a solution of the
chiral model equation (\ref{chir}). Then the action ${\cal{E}}$ defined in
(\ref{ener}) is quantized in integral multiples of $8\pi$}.
\m

As a consequence, the gauge invariant Chern-Simons charge $Q\equiv\int
tr(\Psi^{\dagger} \Psi)$ is quantized in integral multiples of $2\pi \kappa$. A
related quantization condition has been noted in \cite{Dun1}, where the {\it
non-Abelian} charges $Q_\alpha\equiv\int\rho_\alpha$ are quantized in integral
multiples of $2\pi \kappa$ for the $SU(N)$ Toda-type solutions (\ref{sunn}).
(In this case, $Q=\sum_{\alpha}Q_\alpha$).

The relationship between the $SU(2)$ uniton solutions and the $SU(2)$ Toda
solutions illustrated above (\ref{proj}-\ref{diag}) can be generalized to
$SU(N)$ as follows:
\m

\n{\bf THEOREM} \cite{Dun2}: {\it The following matrix}
\beq
h=(-1)^{{1\over 2}N(N+1)} \prod_{\alpha=1}^{N-1} (2p_{\alpha} -1)\
\label{sunh}
\eeq
{\it where $p_\alpha$ is the hermitean holomorphic projector $p_\alpha =
M_{\alpha}(M_{\alpha}^{\dag} M_{\alpha})^{-1}M_{\alpha}^{\dag}$ for the matrix
$M_{\alpha}$ in (\ref{sunn},\ref{uvec}), is a solution of the SU(N) chiral
model equation (\ref{chir}). Furthermore, defining $\chi={1\over
2}h^{-1}\partial_+ h$, there exists a unitary transformation $g$ which
diagonalizes the charge density matrix $[\chi^{\dag}, \chi]$ so that}
\beq
g^{-1}[\chi^{\dag}, \chi] g=\sum_{\alpha =1}^{N-1} \{\partial_+ \partial_- {\rm
log~det}(M_{\alpha}^{\dag} M_{\alpha})\} H_{\alpha}\
\label{sund}
\eeq
{\it where $H_{\alpha}$ are the Cartan subalgebra generators of $SU(N)$ in the
Chevalley basis. This diagonal form is precisely the $SU(N)$ Toda solution
(\ref{sunn}).}

\m

This theorem is proved \cite{Dun2} by expressing the projectors $p_{\alpha}$ in
terms of an orthonormal basis for the space spanned by the columns of $M_{N}$.
The diagonalizing matrix $g$ is also constructed from this orthonormal basis.

\m

In conclusion, I mention some open problems suggested by these results.

\n{1.} The most important physical problem is now to make use of this complete
description of the vacuum of these Chern-Simons-matter theories in order to
develop a second quantized theory.

\n{2.} The fact that this quantization is possible for the $1+1$-dimensional
nonlinear Schr\"odinger equation (NLSE) is intimately related to the
integrability of the $1+1$-dimensional NLSE. Here, in $2+1$-dimensional, the
situation is less clear. Is the $2+1$-dimensional gauged nonlinear
Schr\"odinger equation (\ref{Schr}) with Chern-Simons coupling (\ref{Cher})
integrable?

\n{3.} Can one find time-dependent (i.e. positive energy) solutions other than
those obtained via the action of the dynamical $SO(2,1)$ symmetry acting on the
static solutions?

\n{4.} The work of Uhlenbeck, Wood and Ward gives a beautiful geometrical
picture of the chiral model solutions for the unitary group. What is the {\it
geometrical} interpretation of self-dual Chern-Simons solutions for other Lie
groups? Some solutions, in the Toda form, are known, but the geometrical
understanding of the corresponding chiral model solutions is not clear. This
should be particularly interesting for the self-dual Chern-Simons solutions of
the {\it affine} Toda form.
\b

\n{\bf Acknowledgement:} This work was supported in part by the DOE under grant
DE-FG02-92ER40716.00, and in part by the University of Connecticut Research
Foundation.
 

\begin{thebibliography}{99}

\bibitem{Cang}D.Cangemi, {\it Self-Dual Chern-Simons Solitons with Non-Compact
Groups}, J. Phys. A: Math. and Gen. {\bf 26}, 2945 (1993).
\bibitem{DinZ} A.~Din and W.~Zakrzewski, {\it Properties of General Classical
$CP^{N-1}$ Solutions}, Phys. Lett. {\bf 95B}, 419 (1980); {\it Interpretation
and Further Properties of General Classical $CP^{N-1}$ Solutions}, Nucl. Phys.
{\bf B182}, 151 (1981).
\bibitem{Dun1} G.~Dunne, R.~Jackiw, S-Y.~Pi and C.~Trugenberger, {\it Self-Dual
Chern-Simons Solitons and Two-Dimensional Nonlinear Equations}, Phys.~Rev. {\bf
D43}, 1332 (1991).
\bibitem{Dun2} G.~Dunne, {\it Chern-Simons Solitons, Toda Theories and the
Chiral Model}, Commun. Math. Phys. {\bf 150}, 519 (1992).
\bibitem{Gross} B.~Grossman, {\it Hierarchy of Soliton Solutions to the Gauged
Nonlinear Schr\"odinger Equation on the Plane}, Phys.~Rev.~Lett. {\bf 65}, 3230
(1990).
\bibitem{Hump} See e.g. J.~Humphreys, "Introduction to Lie Algebras and
Representation Theory" (Springer-Verlag 1990).
\bibitem{Jack} R.~Jackiw and S-Y.~Pi, {\it Soliton Solutions to the Gauged
Nonlinear Schr\"odinger Equation on the Plane}, Phys.~Rev.~Lett. {\bf 64}, 2969
(1990); {\it Classical and Quantum Nonrelativistic Chern-Simons Theory},
Phys.~Rev. {\bf D42}, 3500 (1990).
\bibitem{Kost} B.~Kostant, {\it The Solution to a Generalized Toda Lattice and
Representation Theory}, Adv. Math. {\bf 34}, 195 (1979).
\bibitem{Lezn} A.~Leznov and M.~Saveliev, {\it Representation of Zero Curvature
for the System of Nonlinear Partial Differential Equations $x_{\alpha,z
\bar{z}} =exp(kx)_{\alpha}$ and its Integrability}, Lett. Math. Phys. {\bf 3},
389 (1979); {\it Representation Theory and Integration of Nonlinear Spherically
Symmetric Equations of Gauge Theories}, Commun. Math. Phys. {\bf 74}, 111
(1980).
\bibitem{Liou} J.~Liouville, {\it Sur l'\'equation aux diff\'erences
partielles} ${d^2\over du dv}{\rm log} \lambda \pm {\lambda\over 2a^2}=0$,
Journ. Math. Pures Appl. {\bf 18}, 71 (1853).
\bibitem{KUhl} K.~Uhlenbeck, {\it Harmonic Maps into Lie Groups (Classical
Solutions of the Chiral Model)}, preprint (1985), J.~Diff.~Geom. {\bf 30}, 1
(1989).
\bibitem{Vall} G.~Valli, {\it On the Energy Spectrum of Harmonic Two-Spheres in
Unitary Groups}, Topology {\bf 27}, 129 (1988).
\bibitem{Ward} R.~Ward, {\it Classical solutions of the Chiral Model, Unitons
and Holomorphic Vector Bundles}, Commun. Math. Phys. {\bf 128}, 319 (1990).
\bibitem{Wood} J.~C.~Wood, {\it Explicit Construction and Parametrization of
Harmonic Two-Spheres in the Unitary Group}, Proc. Lond. Math. Soc. {\bf 58},
608 (1989).
\bibitem{Zakr} For a review, see W.~Zakrzewski, "Low Dimensional Sigma Models"
(Adam Hilger 1989).

\end{thebibliography}
\end{document}